\documentclass [12pt]{article}
\usepackage [cp1251]{inputenc}
\usepackage [english]{babel}
\usepackage {graphicx}
\usepackage {amssymb}
\usepackage {amsmath}
\usepackage {mathrsfs}
\usepackage {longtable}
\title{Diffusion evolution of a pore in bounded particle in a hydrogen atmosphere}

\author{$^{1}$\textbf{M.I. Kopp},$^{2}$\textbf{P.N. Ostapchuk},$^{1,3}$\textbf{V.V. Yanovsky}}

\begin{document}

\maketitle

$^{1}$ \textit{Institute for single crystals, National Academy of Science of Ukraine, Kharkiv, Ukraine}

$^{2}$ \textit{Institute of Electrophysics and Radiation Technologies, National Academy of Science Ukraine, Kharkov, Ukraine}

$^{3}$\textit{Kharkiv National Kharazin University, Kharkiv, Ukraine}

\abstract{The problem of the diffusion evolution of a pore filled with molecular hydrogen in a spherical granule in a hydrogen medium is solved. The initial position of the pore is displaced relative to the center of the granule. A nonlinear system of equations is obtained, which describes the behavior of the size of the gas-filled pore, the amount of gas in it and its position relative to the center of the bounded particle with time. Numerical calculations have shown the existence of two stages of evolution. The first (fast) stage is associated with the equalization of pressure in the pore with the external. The second is the slow diffusion "healing" of the pore, when the amount of gas adjusts to its size and the gas pressure is approximately equal to the external.}

\section{Introduction}

 The creation of new materials used in nuclear technology, metallurgy, microelectronics, power engineering, instrumentation, space and aviation technology, electrochemical production, solar energy and many other areas is to some extent associated with the problem of porosity, which has a significant impact on the service characteristics of materials. In the general case,  pores have an arbitrary form and size and can be localized both within the elements of the structure of a solid body (for example, inside crystallites, fragments, blocks, cells, or granules) and along their boundaries depending on the prehistory of the substance, its energy balance, and structure \cite{1s}.

Historically, theoretical calculations were initially associated with the so-called diffusion porosity in an unbounded homogeneous medium. According to classical concepts, the diffusion porosity arises in a solid phase supersaturated with point defects due to the migration of excess vacancies and solute gas atoms and includes the stage of pore nucleation and the stage of their growth. As a rule,  a phenomenological approach is used for studying the kinetics of nucleation in multicomponent systems, which is based on the expression for the work $\Delta\Phi\left(\left\{x_i\right\}\right)$  of formation of a new phase nucleus and the Fokker-Planck kinetic equation for the distribution function in the space of variables  $\left\{x_i\right\}$. This approach is an extension of the one-dimensional theory of Zeldovich and Frenkel to the multidimensional case \cite{2s}-\cite{3s}. As a result, the stationary nucleation rate in the space of two or more variables is calculated \cite{4s}-\cite{9s}. The general idea of all works is the reduction of a multidimensional problem to a one-dimensional. In this case, the methods of one-dimensionalization are different, therefore, the pre-exponential factors in the expression for the nucleation rate also are different. As for the growth stage, here the most complete theory has been developed for the so-called coalescence stage. The main contribution to this theory was made by V. V. Slezov and his students \cite{10s}-\cite{17s}.

However, the development of nanotechnologies requires new theoretical researches of defect structures in bounded particles of nano- and meso- scales. The most widespread threedimensional defects in such meso- and nanoparticles are vacancy pores, gas-filled pores as well as new phase inclusions. The regularities of diffusion growth, healing and motion of such defects in nanoparticles is an important problem. Such defect structure plays an important role for the possibility of further compactification of nanoparticles and creating new materials \cite{18s}. Establishing regularities of defect structure evolution will enable one to control it as well as to change properties of corresponding meso- and nanoparticles. The creation of the theory of the diffusive evolution of pores in bounded medias, for example, in spherical nanoparticles, is a rather complicated task. This problem is close to that of diffusion interaction of pores in unbounded matrix \cite{19s}. Indeed, the role of second object the pore interacts with in bounded particles is played by matrix particle boundary. Interaction with boundaries leads to principally different pore behaviour as compared to that in unbounded materials. The formation of pores in spherical nanopaparticles was discovered experimentally in the work \cite{20s}. In the review \cite{21s} the results are presented of theoretical and numerical investigations related to formation and disappearing of pores in spherical and cylindrical nanoparticles. In \cite{21s}, great attention is paid to the problem of hole nanoshell stability, i. e. to the case when in the nanoparticle center large vacancy pores are situated. Analytical theory of diffusive interaction of the nanoshell and the pore situated at arbitrary distance from particle center was considered in the works \cite{22s}-\cite{24s}. With the supposition of quasiequilibrium of diffusive fluxes, the equations have been obtained nonlinear equations for the change of the radii of pore and spherical granule as well as of center-to-center distance between the pore and the granule. It was shown the absence of critical pore size unlike the case of pores in an unbounded matrix \cite{1s}. In the case of a general position, pore in such particles dissolves diffusively, while diminishing in size and shifting towards granule center.

As is known, hydrogen almost always negatively affects the service properties of metals and alloys.
Having high diffusion mobility, hydrogen penetrates into metals to great thicknesses and a macroscale almost evenly over the entire volume of the metal is distributed. The harmful effect of hydrogen extends to the entire volume of the metal. Although the destruction starts from the most structurally dangerous places (stress concentrators, the most stress parts of the product, etc.).

Hydrogen leads to many undesirable changes in the mechanical properties of metals, which hydrogen embrittlement of metals is called. If there is a discontinuity in the material (for example, in the form of a pore), hydrogen actively fills its volume creating high pressure and facilitating the development of cracks.

Naturally, we can consider the problem of the diffusion evolution of a pore filled with molecular hydrogen in a spherical granule in a hydrogen medium
\begin{figure}
  \centering
	\includegraphics[width=13 cm, height=6 cm]{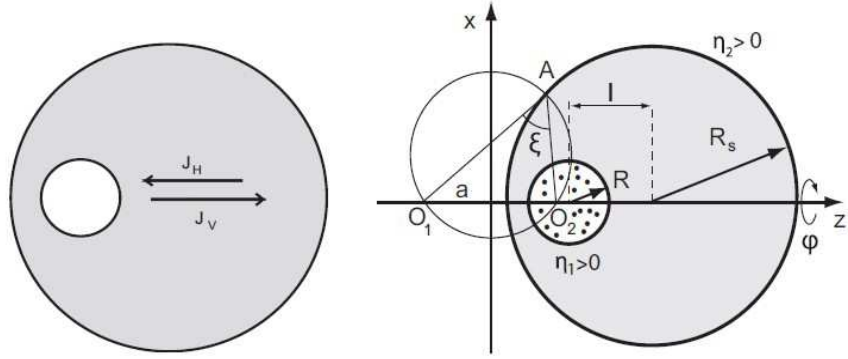}  \\
  \caption{On the left, the filling of a vacancy pore with molecular hydrogen is shown.  There is a one atomic gas (hydrogen) with partial pressure $P_0$ at the granule boundary. On the right,  a gas-filled pore in a bispherical coordinate system is shown. The surface of pore and granule in this coordinate system are coordinate planes $\eta=\textrm{const}$.}  \label{fg1}
\end{figure}

\section{Statement of the problem }

Let us consider the spherical granule of the radius   containing the gas-filled pore of the radius  $R < R_{s}$.
Granule and pore centers are separated from each other by the distance  $l$ (see Fig. \ref{fg1}). The surface of the granule is surrounded by a one component molecular gas (for example, hydrogen) with a constant pressure  $P_0=const$, which fills the pore. Let us assume that the mechanism of pore filling with gas is as follows. The hydrogen molecule $H_2$  breaks up into atoms $H$  on the surface of granule, which penetrate into the matrix forming a solid solution. There are exist various mechanisms of gas atoms diffusion in the crystal lattice of a bounded matrix. The examples of such mechanisms are hopping of matrix atoms, as well as of complexes formed by dopant atoms with vacancies or by other ways. It is assumed that the solution is dilute enough to ignore the interaction of dissolved atoms with each other. Further, hydrogen atoms again collect into gas molecules $H_2$  on the surface of the pore, which fills the pore. In this case, we assume that at the surface of pore and outer boundary of granule in the matrix has been maintained the local thermodynamic equilibrium between the solid solution and molecular gas in the pore and outside granule.

As can be seen from Fig. \ref{fg1}, the geometry of pore and granule boundaries dictates the use of bispherical coordinate system \cite{25s}, as the most convenient one. In bispherical coordinate system  each point $A$  of the space is matched to three numbers  $(\eta,\xi,\varphi)$, where  $\eta=\ln(\frac{|AO_1|}{|AO_2|})$, $\xi=\angle O_1AO_2$, $\varphi$ is polar angle. Let us cite relations connecting bispherical coordinates with Cartesian ones:
 \begin{equation}\label{eq1}
  x=\frac{a\cdot \sin\xi\cdot\cos\varphi}{\cosh\eta-\cos\xi}, \quad
  y=\frac{a\cdot \sin\xi\cdot\sin\varphi}{\cosh\eta-\cos\xi}, \quad
  z=\frac{a\cdot \sinh\eta}{\cosh\eta-\cos\xi},
\end{equation}
where $a$ is the parameter, that at fixed values of pore and granule radii as well as of their center-to-center distance is determined by the relation
$$a=\frac{\sqrt{[(l-R)^2-R_s^2][(l+R)^2-R_s^2]}}{2\cdot l}.$$
Pore and granule surfaces in such coordinate system are given by relations
\begin{equation}\label{eq2}
    \eta_1=\textrm{arsinh} \left(\frac{a}{R}\right), \quad
\eta_2=\textrm{arsinh} \left(\frac{a}{R_s}\right).
\end{equation}
These relations determine values of  $\eta_1$  and  $\eta_2$  from pore and granule radii, while includes additionally center-to center distance $l$  between the pore and the granule.
Thus, the description of diffusion evolution of a gas-filled pore in a bounded particle implies the existence of equations for the rate of pore and granule volume change, the distance between their centers, and the number of gas molecules in the pore with time.
The equations describing the rate of pore and  granule volume change have the form \cite{1s},\cite{26s}:
\begin{equation}\label{eq3}
\dot{R}=-\frac{\omega}{4\pi R^2} \oint\vec{n}\cdot\vec{j}_v|_{\eta=\eta_1} \, dS,  $$
$$\dot{R_s}=-\frac{\omega}{4\pi R_s^2}\oint \vec{n}\cdot\vec{j}_v|_{\eta=\eta_2} dS,  \end{equation}
The rate of changing center-to-center distance between the pore and the granule  is determined by relation
\begin{equation}\label{eq4}
\vec{V}=-\frac{3\omega}{4\pi R^2}\oint \vec{n}(\vec{n}\cdot\vec{j}_v)|_{\eta=\eta_1} dS.
\end{equation}
The integration is carried out over the surface of pore and granule with the outer normal  $\vec{n}$ to them; $\omega$  is the volume per lattice node; $\vec{n}\cdot\vec{j}_v|_{\eta=\eta_{1,2}}$  are the flux densities of vacancies per pore and granule. The pore filling rate with hydrogen has a form similar to (\ref{eq3}):
\begin{equation}\label{eq5}
\dot{N}=-\frac{1}{2} \oint\vec{n}\cdot\vec{j}_H|_{\eta=\eta_1} \, dS,
\end{equation}
where  $\vec{n}\cdot\vec{j}_H|_{\eta=\eta_1}$ is the flux density of atomic hydrogen per pore. These fluxes comply Fick's first law
$$\vec{j}_{v,H}=-\frac {D_{v,H}}{\omega} \nabla C_{v,H} $$
and are determined from the solution of the diffusion problem in the quasi-stationary approximation:
\begin{equation}\label{eq6}
\omega {\rm div}\vec{j}_{v,H}=0
\end{equation}
$C_{v,H}$ is concentration of vacancies and atomic hydrogen in the granule matrix; $D_{v,H}$  are diffusion coefficients.
The approximation (\ref{eq6}) is valid when the characteristic time for establishing the concentration profile is much shorter than characteristic times for changing the pore size and the volume of gas in it. We note an important consequence (\ref{eq6}): the total flow of vacancies through any closed surface is conserved. It means that
$$ \oint\vec{n}\cdot\vec{j}_v|_{\eta=\eta_1}\,dS=\oint\vec{n}\cdot\vec{j}_v|_{\eta=\eta_v}\,dS $$
and the rates of change in the pore and granule volumes are related by $R_s(t)^2\dot{R}_s(t)=R(t)^2\dot{R}(t)$  or:
\begin{equation}\label{eq7}
  R_s(t)^3 ={V+R(t)^3},
\end{equation}
where $V=R_s(0)^3-R(0)^3$ is initial volume of granule material (multiplier $4\pi /3$ is omitted for convenience). The existence  of conservation law (\ref{eq7}) enables us to reduce the number of unknown quantities to three  $R,l$ and $N$.
The expression (\ref{eq6}) takes the form in bispherical coordinates
\begin{equation}\label{eq8}
 \Delta_{\eta,\xi}C_{v,H}=\frac{\partial}{\partial\eta}\left(\frac{1}{\cosh\eta-\cos\xi}\frac{\partial C_{v,H}}{\partial\eta}\right)+
  \frac{1}{\sin\xi}\frac{\partial}{\partial\xi}\left(\frac{\sin\xi}{\cosh\eta-
  \cos\xi}\frac{\partial C_{v,H}}{\partial\xi}\right)= 0
\end{equation}
Here we take into account that due to symmetry of the problem, vacancy concentration does not depend on variable  $\varphi$. The boundary conditions for vacancies are:
$$ C_v(\eta , \xi)|_{\eta_1}=C_{R}^v, \quad C_v(\eta , \xi)|_{\eta_2}=C_{R_s}^v. $$
$C_{R}^v,C_{R_s}^v$  are equilibrium concentrations of vacancies near the spherical surface of pore and granule (see, for example, \cite{1s},\cite{26s}):
\begin{equation}\label{eq9}
C_{R}^v=C_0^{v}\exp\left(\frac{2\gamma\omega}{kTR}-\frac{P \omega}{kT}\right),\; C_{R_{s}}^v=C_0^{v}\exp\left(-\frac{2\gamma\omega}{kTR_s}-\frac{P_0\omega}{kT}\right)
\end{equation}
where  $C_0^{v}$  is equilibrium vacancy concentration near the plane surface, $\gamma$  is surface energy, $T$  is granule temperature, $\omega$ is the volume per lattice site, $P$  is gas pressure inside the pore. For the simplicity we use state equation of ideal gas:
\begin{equation}\label{eq10}
  P\cdot \frac{4\pi}{3}\cdot R^3=N kT
\end{equation}
The boundary conditions for atomic hydrogen are determined by the law of mass action:
\begin{equation}\label{eq11}
  C_R^H = \left(\frac{P \omega}{kT}\delta\right)^{1/2}, \quad C_{R_s}^H=\left(\frac{P_0 \omega}{kT}\delta\right)^{1/2}
\end{equation}
where  $\delta$  is a constant characterizing the thermal  equilibrium at the surface of pore and granule of gas molecules with respect to the chemical reaction of dissociation into constituent atoms
\[\delta = \left(\frac{2\pi h^2}{m \omega^{2/3} k T} \right) \left(\sum_k e^{-\frac{\varepsilon_k}{kT}} \right)^{-1} \cdot e^{- \frac{\psi_H}{kT}}\]
Here  $m$ is the mass of a gas molecule, $h$ - Planck's constant, $\varepsilon_k$ are the energy levels of the resting molecule, the constant $\psi_H$ characterizes the chemical potential of hydrogen atoms for a dilute solid solution $\mu_H = \psi_H + kT \ln C_H $.

Thus, the diffusion problem (\ref{eq8}) is solved once for a "faceless" concentration $C$ with the same "faceless" boundary conditions  $C_R,C_{R_s}$   and then the corresponding rates of change of the sought values are written out taking into account specific expressions for equilibrium concentrations (\ref{eq9})-(\ref{eq11}).

\section{System of equations}

General solution of the equation (\ref{eq8}) with account of boundary conditions is determined as \cite{25s}
\begin{equation}\label{eq12}
C(\eta, \xi ) =
\sqrt{2(\cosh\eta-\cos\xi)}\left\{{C_R}\sum_{k=0}^\infty
\frac{\sinh(k+1/2)(\eta-\eta_2)}{\sinh(k+1/2)(\eta_1-\eta_2)}\exp(-(k+1/2)\eta_1)P_k(\cos\xi)-
\right.$$
 $$ \left.- C_{R_s}\sum_{k=0}^\infty
\frac{\sinh(k+1/2)(\eta-\eta_1)}{\sinh(k+1/2)(\eta_1-\eta_2)}\exp(-(k+1/2)\eta_2)P_k(\cos\xi)
\right\}\,.
\end{equation}
where $P_k$  are the Legendre functions. As a result, for the equations (\ref{eq3})-(\ref{eq5}) we get:
\begin{equation}\label{eq13}
\frac{dR}{dt}=\frac{2D_va}{R^2}\left[ C_{R_s}^v\left(R_s,P_0\right)-C_R^v\left(R,N\right)\right]\Phi\left(\frac{a}{R},\frac{a}{R_s}\right),
\end{equation}
\begin{equation}\label{eq14}
\frac{dN}{dt}=\frac{4\pi D_H a}{\omega}\left[ C_{R_s}^H\left(P_0\right)-C_R^H\left(R,N\right)\right]\Phi\left(\frac{a}{R},\frac{a}{R_s}\right),
\end{equation}
\begin{equation}\label{eq15}
\frac{dl}{dt}=\frac{6D_va^2}{R^3}\left[ C_{R_s}^v\left(R_s,P_0\right)-C_R^v\left(R,N\right)\right]\left\{\Phi\sqrt{1+\frac{R^2}{a^2}}-\widetilde{\Phi}\right\},
\end{equation}
\[\Phi=\sum_{k=0}^\infty \frac{ e^{-(2k+1)\eta_2}}{e^{(2k+1)(\eta_1-\eta_2)}-1}, \quad \widetilde{\Phi}=\sum_{k=0}^\infty \frac{(2k+1)e^{-(2k+1)\eta_2}}{e^{(2k+1)(\eta_1-\eta_2)}-1}. \]
Here it is taken into account that displacement velocity rate along $z$ coincides with $dl/dt$. The details of the calculations are quite cumbersome and included in the Appendix. The equations (\ref{eq9})-(\ref{eq11}), (\ref{eq13})-(\ref{eq15}) and (\ref{eq7}) with the appropriate initial conditions completely determine the evolution of granule and gas-filled pore with time. In the limiting case when the gas is absent and there is no external pressure $P_0=0$, equations (\ref{eq9}), (\ref{eq13}), (\ref{eq15}) coincide with the results of \cite{22s}. The numerical analysis shows that the function $\Phi$ and the expression in braces (\ref{eq15}) are positive over the entire physically reasonable ($R_s>R+l$  pore inside the granule) range of $R_s,R,l$.
Therefore, from (\ref{eq13}) and (\ref{eq15}) immediately follow that the dissolving pore is displaced towards the center of granule and vice versa. Next, we introduce dimensionless variables for convenience:
\[ r = \frac{R}{R_0},\; r_s = \frac{R_{s}}{R_0},\; L = \frac{l}{R_0},\; \alpha=\frac{a}{R_0},\; n=\frac{N}{N_0} ,\; A = \frac{{2\gamma \omega }}{{kTR_0}},\; B=\frac{3\omega N_0}{4\pi R_0^3},\; p_0=\frac{P_0\omega}{kT} , \]
\[t_v=\frac{R_0^2}{DC_0^v}, \quad  t_g=\frac{\omega N_0}{4\pi D_HR_0\sqrt{\delta}},  \quad  \tau =\frac{t}{{t_g }}, \quad q=\frac{t_v}{t_g} .\]
Here $R_0=R|_{t=0}, N_0=N|_{t=0}$ are the radius of pore and the number of gas molecules in at the initial moment of time; $t_v$  and $t_g$  are the characteristic times of change in the size of pore and gas in it.  The complete system of equations (\ref{eq9})-(\ref{eq11}), (\ref{eq13})-(\ref{eq15}), (\ref{eq7}) with initial conditions take the form in new variables:
\begin{equation}\label{eq16}
\begin{cases}
\frac{d r}{d \tau}=\frac{2\alpha}{qr^2}\left[\exp\left(-\frac{A}{r_s}-p_0\right)-\exp\left(\frac{A}{r}-\frac{B}{r^3}\cdot n\right)\right] \cdot  \Phi\left(\frac{\alpha}{r},\frac{\alpha}{r_s}\right),\\
\frac{d n}{d \tau}=\alpha\cdot\left[ \sqrt{p_0}-\sqrt{\frac{B}{r^3}\cdot n}\,\right] \cdot \Phi \left(\frac{\alpha}{r},\frac{\alpha}{r_s}\right),\\
\frac{d L}{d \tau}=\frac{6\alpha^2}{qr^3}\cdot\left[\exp\left(-\frac{A}{r_s}-p_0\right)-\exp\left(\frac{A}{r}-\frac{B}{r^3}\cdot n\right)\right]\cdot \left\{\Phi\sqrt{1+\frac{r^2}{\alpha^2}}-\widetilde{\Phi}\right\},   \\
 r|_{\tau =0}=1, \\
 n|_{\tau =0}=1,  \\
 L|_{\tau =0}=L_0.\\
\end{cases}
\end{equation}
\[ r_s|_{\tau =0}=r_{s0},\; r_s=(r_{s0}^3+r^3-1)^{1/3},\; \alpha = \frac{r_s^2}{2L}\sqrt{1+\left(\frac{L^2}{r_s^2}-\frac{r^2}{r_s^2}\right)^2-2\left(\frac{L^2}{r_s^2}+\frac{r^2}{r_s^2}\right)}. \]
\begin{figure}
  \centering
  \includegraphics[width=16 cm, height=6 cm]{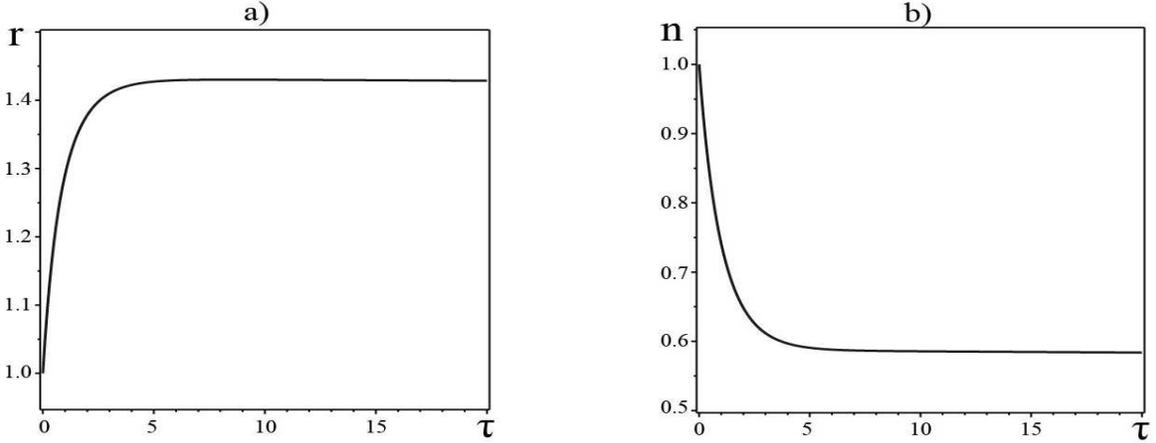}\\
\caption{Plots of the dependence of pore radius $r$  (a) and the number of hydrogen molecules $n$ (b) in the pore on time $\tau$  at $B=2.5 \gg p_0 , r_{s0}=100, L_0=10$.}
\label{fg2}
\end{figure}
Taking into account the equation of state of gas $$p=\frac{n}{r^3}B, $$ we also have $p|_{\tau=0}=B$. For further numerical calculations, we will take: $T=1450^{\circ}\textrm{K}$,  $k=1.38\cdot10^{-16}$ erg/K is Boltzmann constant, $\gamma=10^3$ erg/cm${^2}$ is surface energy density, $\omega=10^{-23}$ cm${^3}$, $R(0)=10^{-4}$cm which corresponds to  $A=10^{-3}$. We take the external gas pressure  $P_0= 10^{10}$dyn/cm${^2}$ or in dimensionless  $p_0=5\cdot10^{-1}$. We set the dimensionless parameter  $q=1$. The values $r_{s0}, L_0, B$  will vary.

\section{Results}

Let consider the case of a "small" pore $(r_{s0} \gg  r_0)$ with initial gas pressure $p|_{\tau=0} \gg p_0$, setting $r_{s0}=100,B=2.5 \gg p_0$. We assume $L_0=10$  (pore far from granule center) for definiteness. Fig. \ref{fg2} shows the dependence of pore radius $r(\tau)$  on time (Fig. \ref{fg2}a) and the change of hydrogen molecules number $n(\tau)$ in the pore on time (Fig. \ref{fg2}b) according to the system of equations (\ref{eq16}).

It can be seen that in a quite short period of time $(\tau\approx 8)$ the pore radius grows and the amount of gas decreases in it reaching some quasi-stationary values $r|_{\tau=8}=1.432$  and $n|_{\tau=8}=0.588$. In this case, the gas pressure in the pore equal to outer pressure $p|_{\tau=8}=p_0$. And then there is a very slow dissolution of the pore but with the condition $p(\tau)\approx p_0$. It follows from system (\ref{eq16}) at $p=p_0 , dn/d\tau=0$. However, the pore radius decreases a little $dr/d\tau <0$  causing an increase of pressure in it and, as a result, a decrease the amount of gas.
There is a gas adjustment to the pore size for maintaining the above condition.

In the other case $p|_{\tau=0} \ll p_0 \, (B=2.5\cdot 10^{-2})$, the process at the initial stage is reversed. The pore radius rapidly decreases in front of increasing the amount of gas and pressure in the pore. The dependences $r(\tau)$  and  $n(\tau)$ on  are shown in Fig. \ref{eq3}. The process stabilizes at reaching pressure $p_0$, and we have a slow dissolution of the pore. The amount of gas is controlled size of pore.
It is clear that in the case $p|_{\tau=0}=p_0 \, (B=0.5)$ we observe immediately regime of slow dissolution of the pore. We note that this behavior of
\begin{figure}
  \centering
  \includegraphics[width=15 cm, height=6 cm]{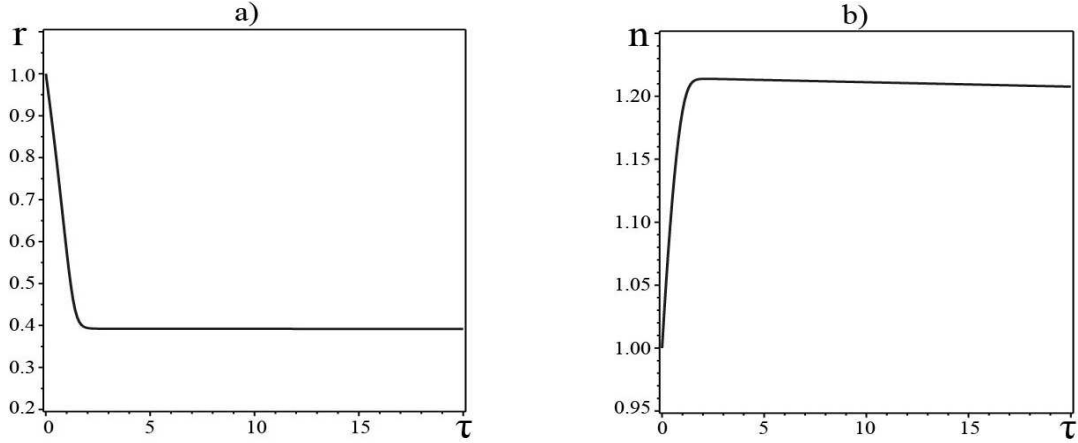}\\
\caption{Plots of the dependence of pore radius $r$  (a) and the number of hydrogen molecules $n$ (b) in the pore on time $\tau$  at $B=2.5\cdot 10^{-2} \ll p_0 , r_{s0}=100, L_0=10$.}
\label{fg3}
\end{figure}
\begin{figure}
  \centering
  \includegraphics[width=18 cm, height=5.5 cm]{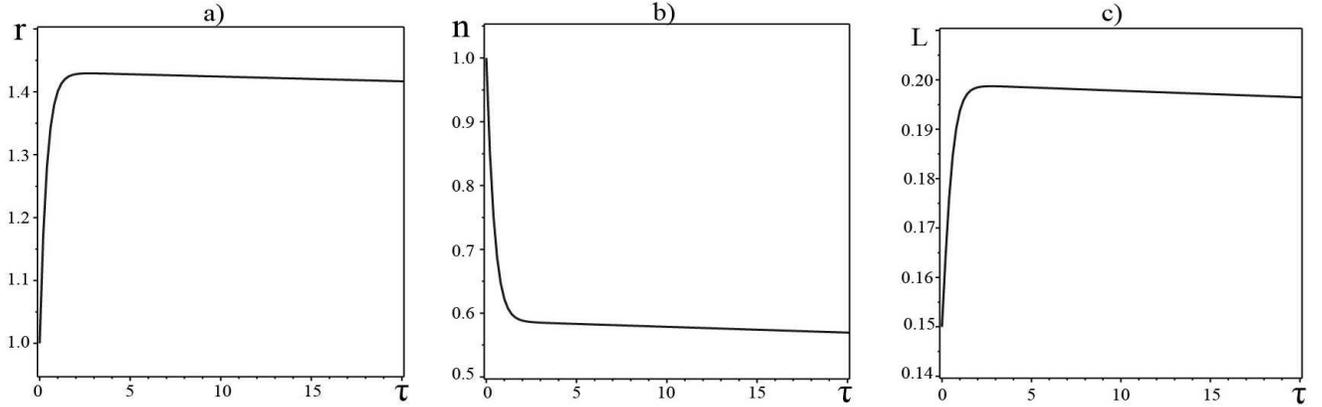}\\
\caption{Plots of the dependence of pore radius $r$ (a) and the number of hydrogen molecules $n$ (b) in the pore on time and the position of pore $L$(c) relative to the center of granule at $B=2.5 \gg p_0, r_{s0}=2,L=0.15$.}
\label{fg4}
\end{figure}
a "small" pore does not depend on its location relative to the center of granule. Similar results were obtained for $L_0=0.1$  (pore near the granule center) and $L_0=60$  (pore near the granule boundary).

The same result is for the "large" pore $(r_{s0}=2>r_0)$.  We must take into account the physical condition $(L+r)<r_s$  (pore inside granule), so we set $L_0=0.15$.
Fig. \ref{fg4} shows the dependence of pore radius $r(\tau)$ (Fig. \ref{fg4}a), the amount of gas $n(\tau)$  in pore (Fig. \ref{fg4}b) on time $\tau$ as well as the position $L(\tau)$  of pore relative to center of granule (Fig. \ref{eq4}c) at $B=2.5 \gg p_0$.

As expected, the pore quickly equalizes the pressure $(p(\tau)\approx p_0)$ and enters the stage of slow "healing". The exit time $(\tau=3)$ is shorter than for the analogous case of a "small" pore, although the situation is the same. We note that the pore is displacement from granule center (see Fig. \ref{fg4}c) at the stage of intensive growth of the pore. However, the above inequality is not violated.

\section{Conclusions}

1. The problem of the diffusion evolution of a pore filled with molecular hydrogen in a spherical granule in a hydrogen medium is solved exactly. Initially, the pore is displaced relative to the center of the granule (see Fig. \ref{fg1}). \\
2. A nonlinear system of equations (\ref{eq16}) is obtained, which describes the time behavior of the size of a gas-filled pore, the amount of gas in the pore, and the position of the pore relative to the center of a bounded particle. \\
3. The system (\ref{eq16}) was solved numerically. The calculations showed the presence of two stages of pore evolution. The first (fast) stage is associated with the equalization of pressure in the pore with the external one. The second stage is the slow diffusion "healing" of  pore when the amount of gas adjusts to its size, and the gas pressure is approximately equal to the external one.

\end{document}